# Ion Track Formation via Electric-Field-Enhanced Energy Deposition

Zikang Ge,[1,#] Jinhao Hu,[1,#] Shengyuan Peng,[1] Wei Kang,[2] Xiaofei Shen,[3] Yanbo Xie,[4,*] and Jianming Xue[1,2,†]

[1]*State Key Laboratory of Nuclear Physics and Technology, School of Physics, Peking University, Beijing 100871, China*

[2]*CAPT and HEDPS, College of Engineering, Peking University, Beijing 100871, China*

[3]*Center for Applied Physics and Technology, HEDPS, and SKLNPT, School of Physics, Peking University, Beijing 100871, China*

[4] *Institute of Extreme Mechanics, School of Aeronautics, Northwestern Polytechnical University, Xi'an 710072, China*

High-energy ion irradiation deposits extreme energy in a narrow range (~1-10 nm) along ion trajectories in solid through electronic energy loss, producing unique irradiation effects such as ion tracks. However, intrinsic velocity effects impose an upper limit on electronic energy loss that cannot be overcome by adjusting irradiation parameters. We introduce a method using electric fields during irradiation to enhance nanoscale energy deposition by accelerating ion-excited electrons within sub-picosecond timescales. Our extended field-enhanced thermal spike model quantitatively describes this enhancement and predicts a significant reduction in the electronic energy loss required for ion track formation in amorphous $SiO_2$, which is in excellent agreement with experimental observations. This work provides a new approach to control energy deposition during irradiation and boosts the wide application of ion tracks in material modification and nanoengineering to much broader extents.

## I. Introduction

When an energetic ion penetrates a solid, its kinetic energy loss will transfer to the electronic system through inelastic interactions (electronic stopping, $S_e$) and to the atomic system via elastic collisions (nuclear stopping, $S_n$). For high-energy irradiation, energy deposition in the target solid is dominated by $S_e$, while $S_n$ is negligible[1, 2]. This energy deposition induces rapid local heating in the solid on tens nanometer spatial scales and sub-picosecond timescales via electro-phonon coupling, followed by heat dissipation over tens to hundreds of picoseconds, forming a thermal spike that brings about various irradiation effects, such as plastic deformation[3], ionization-induced defect annealing[4, 5], or even narrow trails of permanent damage along the ion path[6], called "ion track"[2, 7]. These unique effects have been extensively used to characterize and modify materials, leading to broad applications in microelectronics[8], geological dating[9], biotechnology[10], and nanomaterial engineering[11-13].

Controlling the confined nanoscale energy deposition $E_c$ in solids is the basis of generating and applying these effects[14]. Generally, $E_c$ equals $S_e$ and can be modulated by varying the incident ion type and energy when using an accelerator for high-energy ion irradiation. However, due to intrinsic velocity effects[15], $S_e$ values exhibit upper limits that cannot be exceeded by merely adjusting ion parameters. This imposes significant constraints on irradiation applications. For example, ion track formation remains challenging in diamond[16] and crystalline silicon[17], or even impossible in certain materials like silicon carbide[18]. Therefore, in addition to modifying ion parameters, developing an alternative method to enhance the confined energy deposition is essential.

In this study, we introduce a method using external electric fields during irradiation to enhance $E_c$, which enables ion-excited electrons to gain additional energy within sub-picoseconds. Experimental results in this work demonstrate that the applied electric field during irradiation can markedly enhance nanoscale energy deposition, thereby promoting ion track formation in amorphous $SiO_2$ (a-$SiO_2$) even under low $S_e$ irradiation. Meanwhile, we extended the thermal spike model and employed multi-physics finite element simulations to describe the field-enhanced $E_c$ (comprising both ion-induced electronic energy loss and the field-accelerated electron energy) and subsequent thermal spike effect. The simulations indicate that field-induced electron energy synergistically evolves with electronic energy loss during sub-picosecond electron-phonon coupling, intensifying thermal spike effects and raising the temperature above the ion track formation threshold, in good agreement with our experiments. This method may open up new research areas for irradiation effects and enable innovative applications including ion-track-based nanopatterning[19, 20], defect engineering[21, 22], and targeted material modification at the nanoscale[23, 24].

* Contact author: ybxie@nwpu.edu.cn

† Contact author: jmxue@pku.edu.cn

# II. Experiment and Results

## A. Experimental setup

We first demonstrate through a carefully designed ion track formation experiment in a-$SiO_2$ that energy deposition is substantially enhanced when an electric field is applied during ion irradiation. As a key material for ion-track-based nanofabrication, a-$SiO_2$ has been used as a model system for elucidating the mechanism of ion track formation over the past decades[2, 25].

Ion track formation in solids evolves with increasing nanoscale energy deposition from no track to discontinuous tracks and continuous tracks, leading to distinct morphologies during characterization (e.g., SEM after etching)[26-28]. Therefore, we use a comparison of ion track formation with and without an applied electric field as the criterion for field-enhanced nanoscale energy deposition.

In our experiments, 280 nm thick a-$SiO_2$ thin films thermally grown on Si substrates are used as samples. These samples are divided into two experimental groups: one subjected to conventional irradiation alone, and another to our electric field enhancement method. We deposited an electron-beam-evaporated Au/Ti layer on the sample surface to serve as an electrode as shown in Figure 1, and all irradiated samples were equipped with this Au/Ti layer, including the control group.

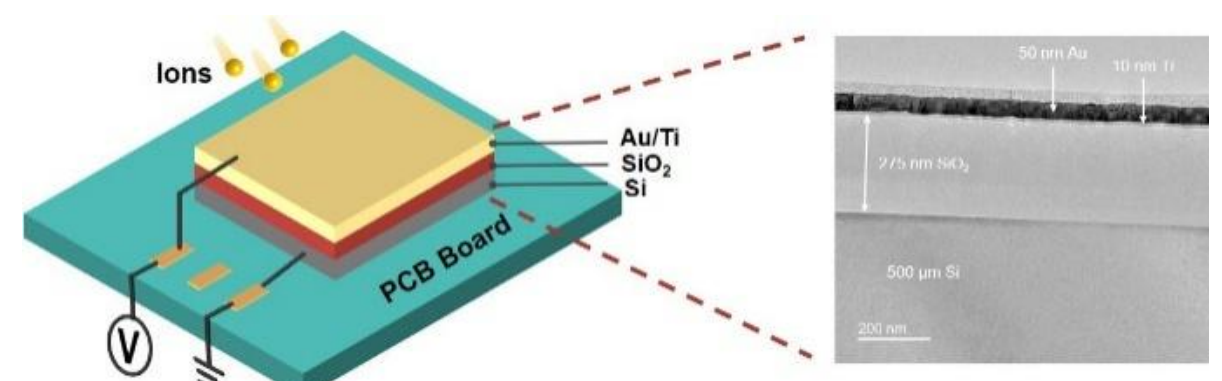

**FIG. 1** (color online)**.** Schematic of the irradiation test samples. An Electron-beam-evaporated Au/Ti top layer and a heavily doped Si substrate serve as electrodes.

The two groups of samples were irradiated at a 2×1.7 MV and 2 × 6 MV tandem accelerator at Peking University, with 2.6 MeV C ions, 4.1 MeV F ions, 6 MeV Si ions, and 20 MeV Si ions at room temperature in a vacuum. Ion impact densities were $2 \times 10^9$ to $1 \times 10^{10}$ ions/$cm^2$, and the angles of incidence were set to be 0° with respect to the surface normal. What's different is that the first group was directly irradiated without bias, while the second was irradiated under a uniform electric field—supplied by a DC voltage and aligned with the ion incidence direction—as schematically shown in Figure 1. Under these irradiation conditions, $S_e$ remains below the *continuous* track threshold in a-$SiO_2$ ($S_e < S_e^{th} = E_c^{th}$=4 keV/nm[28, 29], Table I), theoretically unable to form continuous ion tracks and likely limited to discontinuous damage features. In fact, due to the velocity effect, the maximum $S_e$ values for C, F, and Si ions irradiation all remain below $S_e^{th}$ regardless of the incident energy adjustments, as shown in Figure 2. This suggests that, under conventional irradiation conditions, these ions are unlikely to generate continuous ion tracks in a-$SiO_2$ and at most may produce discontinuous damage features. Notably, as long as ion track formation is significantly enhanced in the second group compared to the first group, we can conclusively establish that external electric fields significantly enhance energy deposition.

TABLE I. Ion irradiation parameters. The $S_e$ and $S_n$ are calculated by SRIM-2013 package[30]

| Ion | $E$（MeV） | $S_e$（keV/nm） | $S_n$（keV/nm） |
|---|---|---|---|
| C | 2.6 | 1.26 | 0.004 |
| F | 4.1 | 2.01 | 0.009 |
| Si | 6.0 | 2.85 | 0.020 |
| Si | 20 | 3.48 | 0.007 |

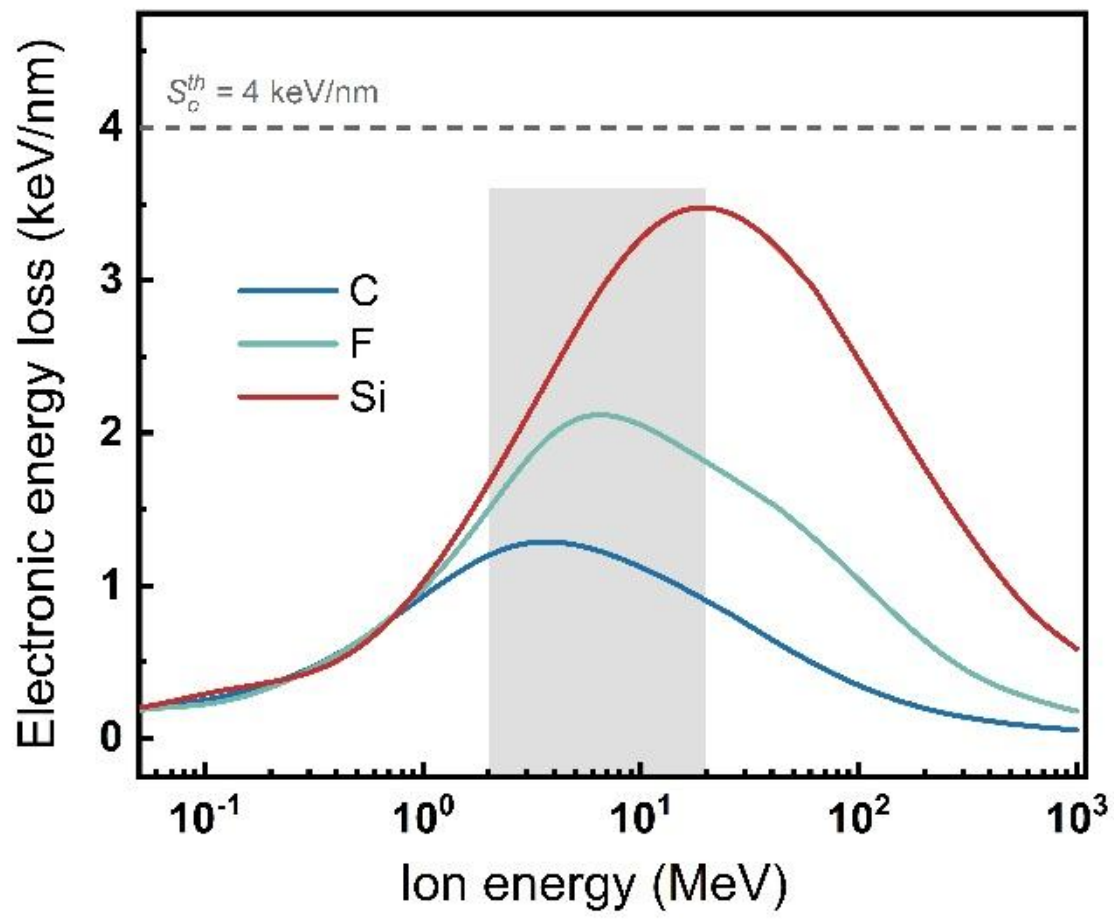

**FIG. 2** (color online)**.** Electronic energy loss as a function of ion energy for three types of ions (C, F, Si) incidence a-$SiO_2$, calculated by SRIM-2013 package[30]. The shaded area indicates the ion energy range used in this irradiation experiment. The dashed line represents the energy deposition threshold required for ion track formation in a-$SiO_2$.

Since ion tracks in a-$SiO_2$ are 'latent tracks' difficult to observe directly, we employed wet etching to assess their formation. In principle, tracks become discernible when the local etch rate along the track ($v_t$) exceeds the bulk etch rate ($v_b$). Accordingly, samples with ion track formation exhibit conspicuous surface etched pits after etching [26-28]. In our experiments, the irradiated samples were mechanically stripped of the surface Au/Ti layer using lift-off tape, followed by ultrasonic cleaning in ethanol and deionized water. Then the irradiated samples were etched in a 4% hydrofluoric acid (HF) aqueous solution at room temperature for 5 minutes[28] to reveal tracks, then rinsed with deionized water, dried with nitrogen, and characterized by SEM[31].

## B. Experimental results

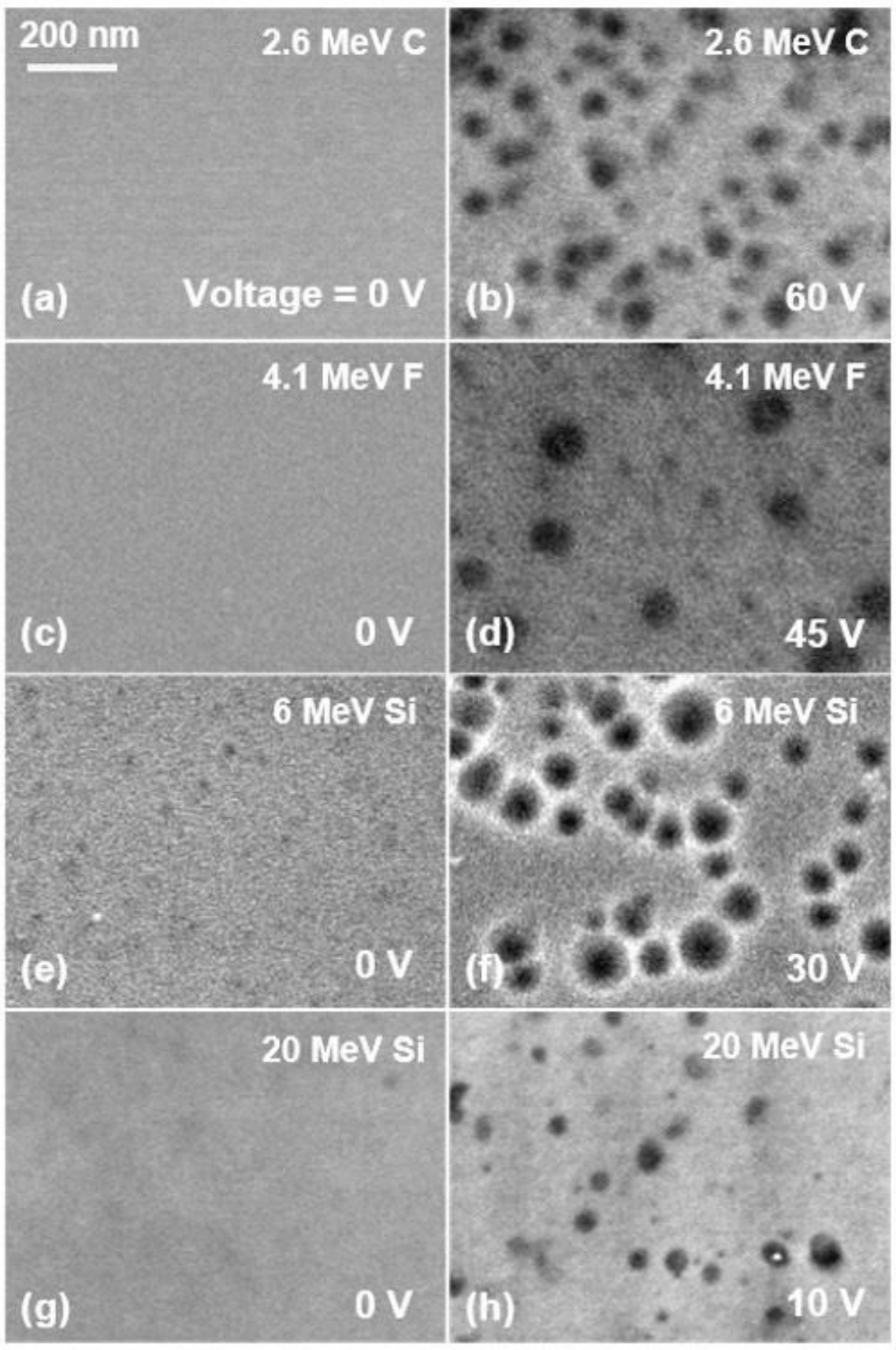

**FIG. 3.** SEM images of a-$SiO_2$ films under no electric field and applied field irradiated by (a)-(b) 2.6 MeV C ions, (c)-(d) 4.1 MeV F ions, (e)-(f) 6 MeV Si ions and (g)-(h) 20 MeV Si ions with $2 \times 10^9$ to $1 \times 10^{10}$ ions/cm$^2$. The voltage applied during the irradiation is given in each image, and the electric field strength $E$ equals the ratio of voltage to thickness 280 nm, $E$ = voltage/280 nm.

Figure 3 compares the SEM surface morphologies of a-$SiO_2$ under conventional (field-free) irradiation conditions and electric field-enhanced conditions. Figs. 3(a), (c), (e), and (g) display the etched surfaces irradiated with the conventional method; no obvious etched pits of ion tracks (typically appearing as round black spots in the images) can be observed on the a-$SiO_2$ surfaces. This aligns with general expectations: under these conditions, most tracks do not form to be etched or are difficult to clearly discern. However, as shown in Figs. 3(b), (d), (f), and (h), etched pits of ion tracks become clearly visible when a strong enough electric field is applied, in sharp contrast to the field-free case. This provides direct experimental evidence of the enhancement effect of the applied electric field on the ion track formation. Further SEM results are available in Figure 4.

Notably, due to differences in irradiating ion species, sample preparation, and etching conditions, the reported $S_e$ threshold for etchable tracks in a-$SiO_2$ varies across the literature, spanning approximately 1.5–4 keV/nm[28, 29]. For the samples irradiated with 20 MeV Si ($S_e$=3.48 keV/nm) in this work, most ion impacts did not result in etchable tracks, fully consistent with the experimental observations reported in Ref. [29].

Accordingly, we adopt the widely accepted threshold for *continuous* track formation ($S_e^{th}$ = 4 keV/nm) as our working criterion, which also renders the with/without-field comparison clearer and more transparent.

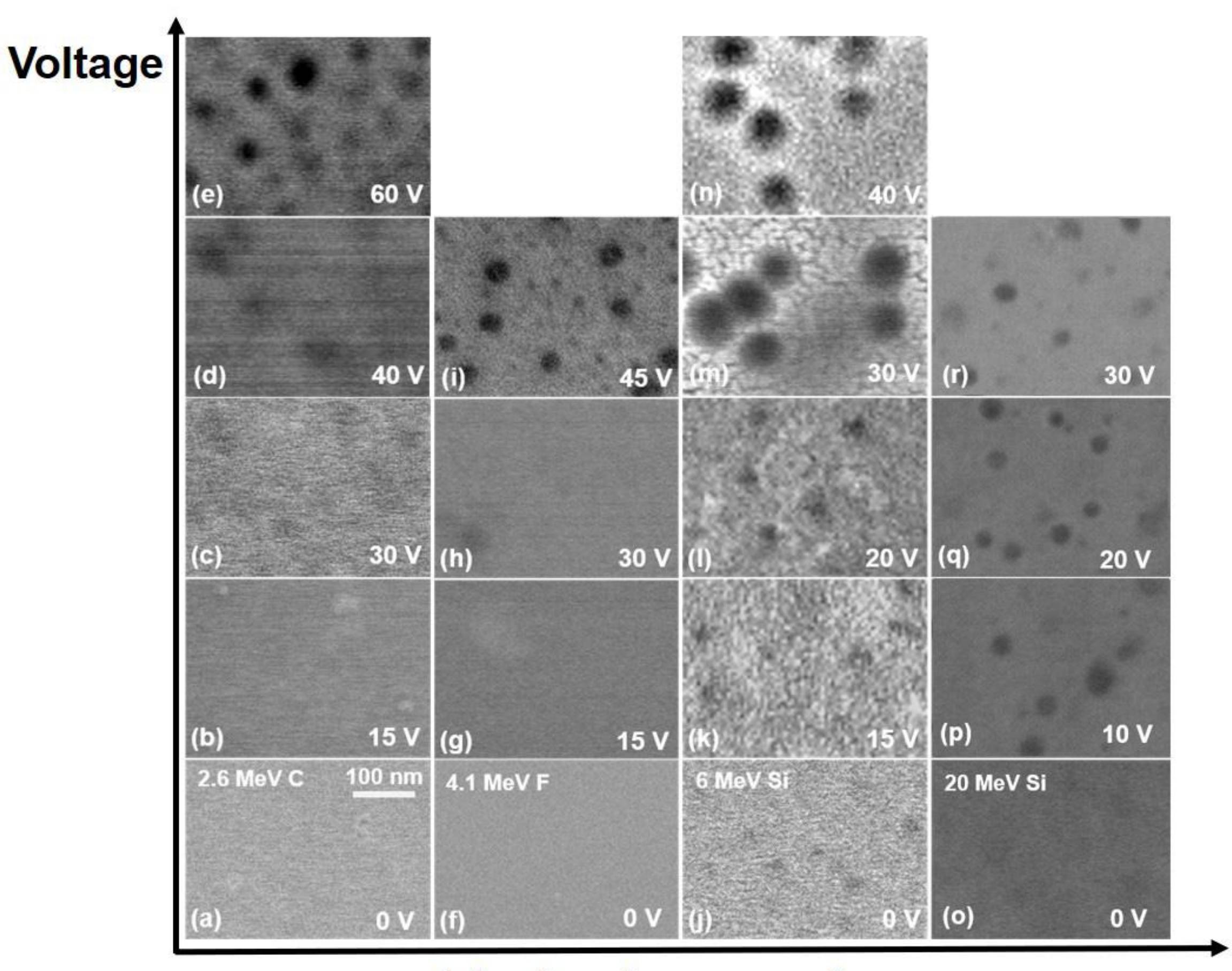

**FIG. 4.** SEM images of a-$SiO_2$ films subjected to different voltage irradiated by (a)-(e) 2.6 MeV C ions, (f)-(i) 4.1 MeV F ions, (j)-(n) 6 MeV Si and (o)-(r) 20 MeV Si. The voltage applied across both film surfaces is given in each image.

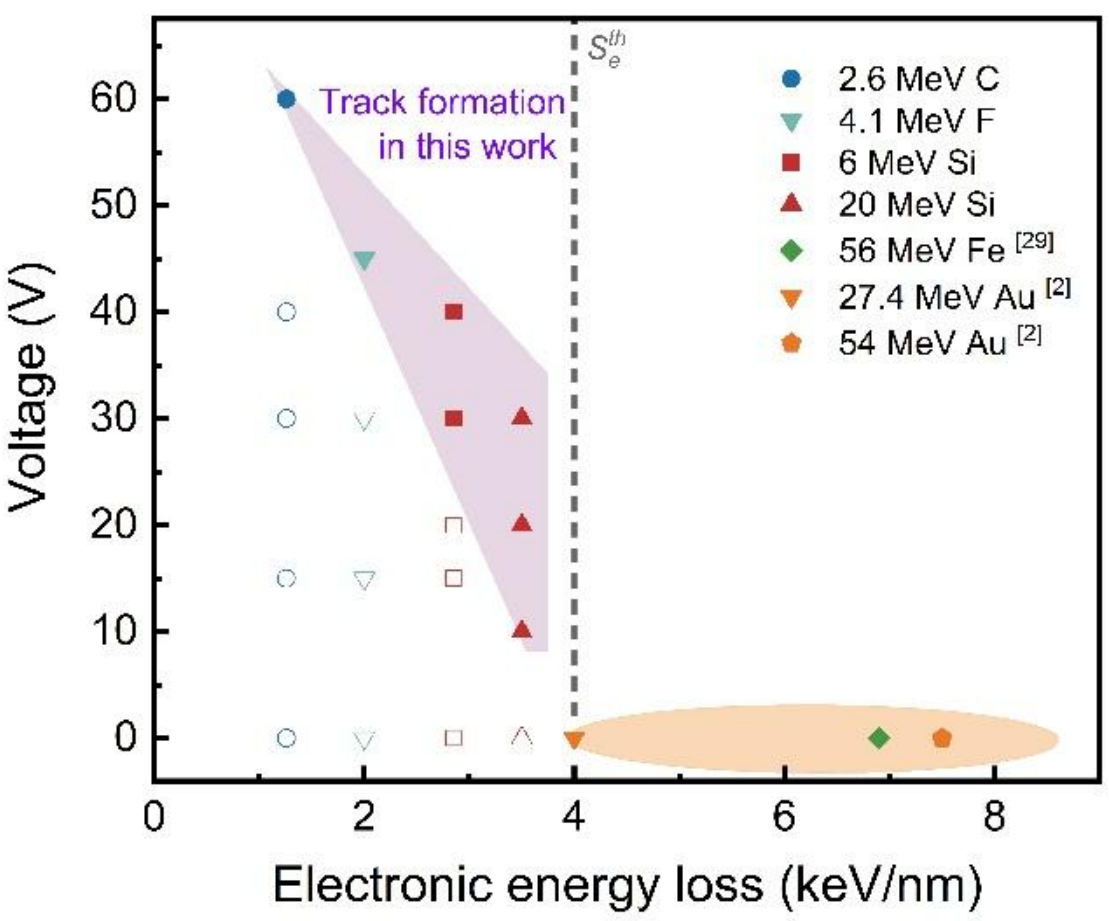

**FIG. 5** (color online). Experimental results of ion track formation in a-$SiO_2$ (points) under different electronic energy loss and electric field conditions. Solid points indicate ion track formation, while hollow points indicate no formation.

Figure 4 presents the ion track formation for four irradiation conditions with different $S_e$ under various applied electric fields. As $S_e$ decreases, a larger field-induced energy deposition is required to form ion tracks, and thus stronger external fields are needed—a physically natural outcome.

Figure 5 displays the ion track formation conditions using the electric field enhancement method (purple shaded area), which is located in the low-$S_e$ irradiation region below the ion track formation threshold $S_e^{th}$ of a-$SiO_2$ (shown by the vertical dashed line in the middle), also included is the high-$S_e$ irradiation region (orange shaded area) using the conventional irradiation method. It shows that under irradiation conditions below the conventional $S_e^{th}$, the new method is still able to successfully form ion tracks, and the required external electric field is roughly a diagonal line in the left part, which is just the threshold energy $E_c^{th}$. Ion tracks cannot form in the low-$E_c$ irradiation region (white area in the bottom left). These results confirm that applying an external electric field during irradiation significantly enhances nanoscale energy deposition and effectively promotes ion-track formation even at low $S_e$.

# III. Method and Simulation

## A. Field-enhanced thermal spike model

We now discuss the underlying physical mechanisms responsible for electric-field–enhanced ion track formation. The thermal spike model is widely regarded as well describing the nanoscale energy deposition and ion track formation in insulators[32-35]. These models propose that local electrons initially acquire energy from incident ions, followed by energy transfer to lattice atoms through electron-phonon coupling within an extremely short timeframe (<1 ps)[23], generating localized high-temperature thermal spikes. When the lattice temperature near the ion trajectory rises above the target material's melting point, a melting-phase transition occurs along the trajectory, producing a continuous, heavily damaged zone and thereby giving rise to ion-track formation.

In conventional thermal spike models (e.g., the Analytic Thermal Spike Model, ATSM[34]), the heat transported through electron–phonon coupling arises exclusively from the electronic energy loss, with no additional sources considered. According to the ATSM, the temperature distribution within the target materials resulting from the SHIs exhibits a Gaussian profile[34, 36], which can be written as:

$$T(r,0)=\frac{Q}{\pi a(0)^2\rho c}e^{-\frac{r^2}{a(0)^2}}$$

Here, $r$ represents the radial distance from ion path, $a(0)$ the standard deviation of the Gaussian distribution, $\rho$ the density, and c the heat capacity of the target material. $Q$ is the energy absorbed by the localized lattice system per unit depth of target from the electronic system. In the conventional thermal spike model, $Q$ was entirely originating from $S_e$ caused by irradiation, written as $Q = gS_e$, where $g$ is the electron-phonon coupling parameter, indicating the fraction of electronic energy loss contributing to heating. This formula clearly reveals the relationship between temperature and $S_e$. Because ion-track formation requires the temperature to reach the target material's melting point, it also explains why a threshold $S_e^{th}$ must be attained in specific materials.

However, in this work, the energy deposition has gained a new source from external electric fields, and the subsequent thermal evolution process has also changed, requiring us to extend the field-enhanced thermal spike model.

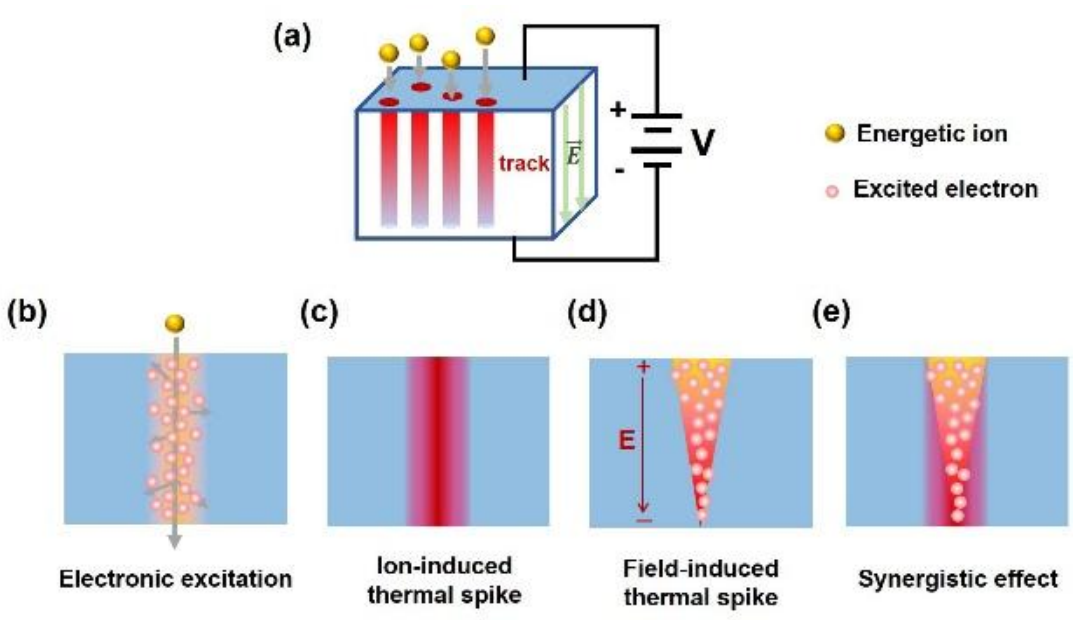

**FIG. 6** (color online)**.** Schematic diagram of the field-enhanced thermal spike model:(a) track formation in the sample irradiated by energetic ions under electric fields, (b) electronic excitation induced by electronic energy loss, (c) thermal spike generated by ion-induced electronic energy loss, (d) thermal spike generated by field-accelerated electron energy, and (e) synergistic thermal spike effect arising from the coupling (c) and (d).

Figure 6 illustrates our extended field-enhanced thermal spike model incorporating electric field-irradiation synergistic effects. Upon initial ion incidence, electronic energy deposition induces localized electronic excitation within 10 fs[23], generating abundant electrons. The number of these electrons per unit depth can be quantified as:

$$n = \frac{S_e}{\omega} = \frac{S_e}{2.73E_g + 0.55(\mathrm{eV})} \text{ [37]}$$

where $E_g$ represents the bandgap energy of 8.9 eV characteristic of a-$SiO_2$[38].

In the presence of an external electric field, the excited electrons are directionally accelerated parallel to the field, giving rise to a transient current density $J$, which is determined by the combined effects of the applied electric field $E$ and electron concentration $n$. The ensuing charge-transport dynamics are governed by the following electrical equations:

$$\nabla \cdot (\varepsilon \nabla V) = -\rho_c$$

$$\frac{\partial n}{\partial t} + \frac{1}{q}\nabla \cdot J = 0$$

$$J = q\mu_n n(-\nabla V) + qD_n \nabla n$$

In this way, electrons can acquire additional energy from the electric field beyond that provided by electronic energy loss. Under a strong field, the drift current formed by electrons parallel to the field is confined around the ion track. These excited electrons thereby gain significant field-induced kinetic energy, intensifying electron-lattice collisions and amplifying the thermal spike effect. The power density per unit depth $P$ is determined by the current density $J$ and the electric field strength $E$, written as:

$$P = \int J \cdot E \, dS$$

The ATSM neglects the electron–phonon coupling process by setting the end of electron–phonon coupling at t=0 and directly prescribing the temperature distribution in the target. This simplification is insufficient for our field-enhanced thermal spike model, where the transfer of heat from ion-induced electronic energy loss and field-induced electron energy to the atomic system must be accounted for. Accordingly, we define the initial time (t=0) as the moment that the incident ion exits the sample and its energy loss has been fully deposited into the solid's electronic system, and then initiate the calculation of the heat-transfer process. We extended the conventional thermal spike model by modifying the electron-phonon coupling energy $Q$ to a field-enhanced heat $Q^*$ that encompasses all energy transfer processes under electric field-enhanced conditions: The following thermal equations govern this process under the synergistic effects of irradiation and applied electric field:

$$Q^* = Q(S_e) + Q(J) = gS_e + \underbrace{\iint J \cdot E dSdt}_{\text{electric field-induced term}}$$

$$\rho C_p \frac{\partial T}{\partial t} = \nabla \cdot (\kappa \nabla T) + \frac{\partial^2 Q^*}{\partial S \partial t}$$

The Fourier heat conduction equation governs the relationship between the target material's temperature T and the heat transfer $Q^*$. In insulators (e.g., a-$SiO_2$), the thermal conductivity $\kappa$ is small, yielding strong spatial localization of deposited heat and limited lateral dissipation; consequently, high temperatures readily develop around the ion track. Moreover, insulators generally possess weak bonding and low melting points, which facilitate collective bond breaking and phase transformation at elevated temperatures. In contrast, semiconductors such as Si and SiC exhibit larger $\kappa$, stronger bonding, and higher melting points, rendering melt-mediated phase transitions far less probable; metals accentuate these trends. These trends explain why the $S_e$ required for ion track formation increases from insulators to semiconductors to metals.

The intricate nature of electron dynamics presents significant challenges in accurately modeling electron-phonon coupling processes. To address this complexity, we implemented a pragmatic approach that decouples the calculation of electronic energy loss and field-induced energy contributions, treating these energy deposition mechanisms as independent phenomena. In our model, the $S_e$ component is uniformly fed into the system over a characteristic timescale of 150 fs for a-$SiO_2$[39], incorporating an electron-phonon coupling coefficient $g = 0.4$[34, 35], to quantify the energy transfer efficiency. Concurrently, the temporal extent of field-induced energy deposition is dynamically determined by electron transport velocities, with the process naturally terminating upon electron migration to boundaries where recombination occurs. Although this methodological separation of electron energy contributions represents a simplification of the actual physics, it renders the synergistic interaction between ion-induced $S_e$ and field-induced electron energy more transparent, as schematically depicted in Figure 6. Meanwhile, the agreement between our simulations and experiments further supports the validity and practical utility of this simplified modeling approach.

To quantitatively illustrate the subsequent effects of electric field-enhanced energy deposition, we construct a finite element simulation based on the extended field-enhanced thermal spike model using COMSOL Multiphysics software. This simulation focuses on the temperature response of a-$SiO_2$ to assess whether the field-enhanced energy deposition method can promote ion track formation, and numerically reveals the thermal spike effect under varying bias voltages (electric-field intensities) and $S_e$ conditions. Our simulation, referencing the ATSM, assumes that the irradiation-induced electronic energy deposition and the instantaneous electronic excitation follow a Gaussian profile with a(0) = 4.5 nm in a-$SiO_2$[35]. We take $T_m$ = 2100 K (with a phase transition latent heat of 1.14 eV/molecule[2]) as the melting temperature of a-$SiO_2$, which serves as the criterion for continuous ion-track formation. Under zero electric field, the calculated track temperature attains this threshold $T_m$ when $S_e$ = 4 keV/nm.

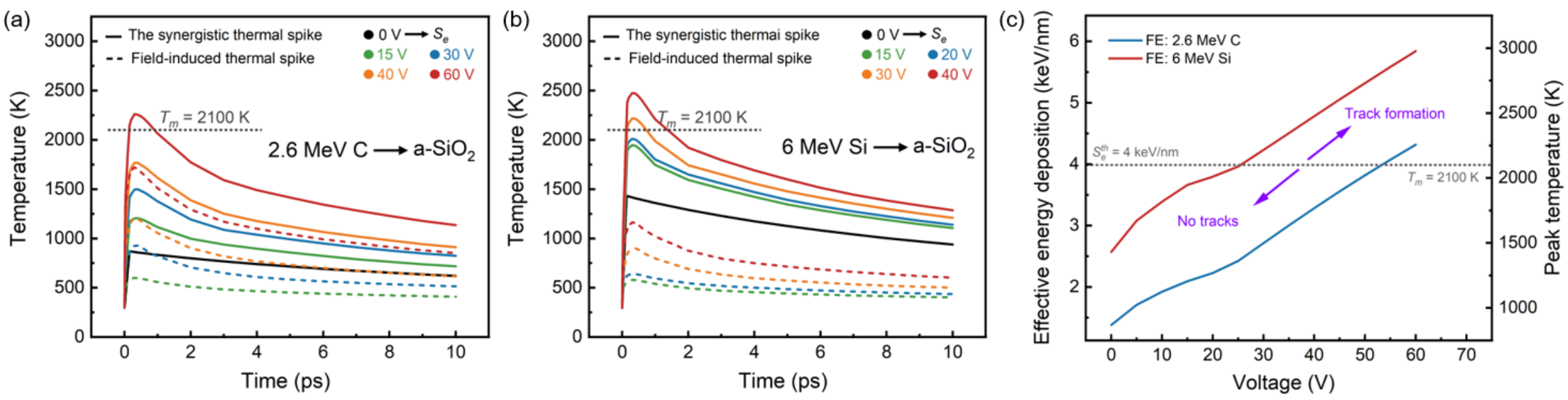

**FIG. 7** (color online). Field-enhanced thermal spike effect in a-$SiO_2$. (a,b) Simulated temperature evolution induced by 2.6 MeV C and 6 MeV Si ions, respectively. Solid lines show synergistic thermal spikes from combined ion and field effects; black solid lines represent conventional free-field cases ($S_e$ only); dashed lines show thermal spikes from field effects alone. (c) Effective energy deposition (left axis) and peak temperature (right axis) as functions of applied voltage. Gray dotted lines in all panels denote ion track formation thresholds. All results were obtained from FE simulations incorporating the field-enhanced thermal spike model.

## B. Simulation results

The synergistic effect of thermal spikes induced by both incident ions and applied electric fields is key to promoting ion track formation. We performed FE simulations with varying electric field strengths under 2.6 MeV C and 6 MeV Si ion irradiation conditions to reveal this enhanced thermal spike effect, with results shown in Figure 7(a) and (b). For the same given irradiation conditions, the contribution of ion-induced electron energy deposition (i.e., electronic energy loss, $S_e$) to the thermal spike remains fixed (black solid lines in Figure 7, representing thermal spikes without electric field). The $S_e$-driven thermal spike increases the target temperature during the 150 fs electron–phonon coupling period; once the electronic energy transfer ceases, the temperature declines. The heat transport follows the Fourier heat conduction equation. While the field-induced electron energy increases with external field strength, it subsequently enhances the thermal spike effect. To determine the contribution of field-induced electron energy to the synergistic thermal spike, we also simulate thermal spikes formed solely by field-induced electron energy (dashed lines in Figure 7). The results show that field-induced electron energy is sufficiently strong and field-induced thermal spike effects are comparable to or even exceed those of ion-induced thermal spikes. Field-induced electron-lattice energy transfer is complete within 0.3 ps, nearly in sync with the ion-induced thermal spike, enabling a strong synergy between the two energy deposition components. Simultaneously, within such short timescale, electrons propagate only along the ion trajectory (parallel to the electric field $E$), with radial diffusion suppressed to a negligible level. Consequently, both energy depositions occur within the same spatial domain. This explains why the electric field enhancement method significantly enhances the confined energy deposition.

Under identical irradiation conditions, leveraging the field–ion synergistic effect, increasing the external electric field strengthens field-induced electronic energy deposition, thereby continuously raising the material's peak temperature. Based on their proportional contribution to thermal spikes, field-induced electron energy can be equivalently treated as an *effective electronic energy loss*, facilitating direct comparison with effective energy deposition $E_c$. When energy deposition exceeds the threshold, the local temperature of a-$SiO_2$ will simultaneously exceed the phase transition temperature, successfully forming ion tracks. Figure 7(c) illustrates how energy deposition and peak temperature vary with external electric field under 2.6 MeV C and 6 MeV Si ion irradiation, readily revealing the electric field thresholds required for ion track formation under low-$S_e$ irradiation conditions. Lower $S_e$ means more supplemental energy and fewer excited electrons, thus a stronger external field is required.

The new $E_c^{th}$ conditions for ion track formation under electric fields obtained from FE simulation are shown in Figure 8, agreeing extremely well with the experimental data in Figure 5. With the electric field-enhanced method, this threshold curve replaces the conventional irradiation $S_e^{th}$, revealing the pattern of electric field-enhanced energy deposition, and providing new insights for ion track formation in various solids.

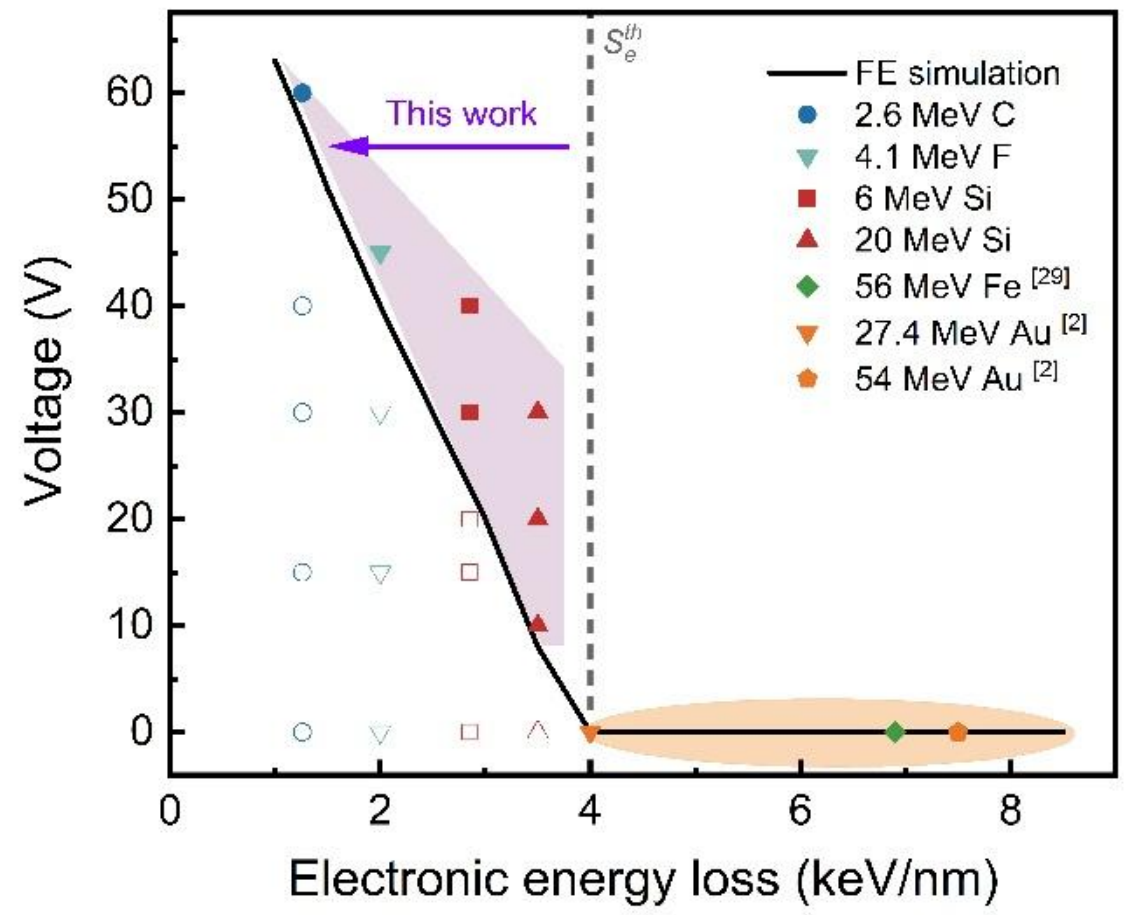

**FIG. 8** (color online)**.** The new track formation criteria (black solid line) obtained from the FE simulation. Solid points indicate ion track formation, while hollow points indicate no formation.

Semiconductor and metallic materials rarely form ion tracks under conventional irradiation, limiting their applications in fields such as nanoscale circuit patterning[40]. However, their superior electron mobility $\mu_n$ compared to a-$SiO_2$ facilitates stronger and more concentrated energy deposition in electric fields. Theoretically, semiconductors and metals could exhibit better electric field regulation effects, forming ion tracks under this new method. Additionally, the electric field enhancement method can be extended to other irradiation applications, such as ionization-induced annealing of pre-existing defects in various materials[4], offering broad applicability.

# IV. Conclusion

In conclusion, this work establishes a new method that applies external electric fields during irradiation to enhance nanoscale energy deposition. This method enhances electron-phonon coupling in the critical sub-picosecond timeframe, enabling control over effects that were previously inaccessible at given irradiation parameters. The new field-enhanced method demonstrably promotes ion-track formation in a-$SiO_2$ under low-$S_e$ irradiation by C, F, and Si ions—species for which track formation in a-$SiO_2$ has not been previously reported—thereby confirming the method's effectiveness.

We extended the field-enhanced thermal spike model by combining ion-induced electronic energy loss with field-induced Joule heating to elucidate the physical mechanism of this method. Based on this extended model, we first simulated the thermal spike process under electric-field enhancement and then obtained a new threshold curve via multiphysics finite-element simulations, which shows excellent agreement with the experimental results. While demonstrated by ion track formation in a-$SiO_2$, further studies across different materials and irradiation regimes will be essential to fully establish its potential.

Overall, this work opens a new research area and fills a theoretical gap in ion-matter interaction. The electric-field-enhanced irradiation method has practical significance and potential applicability for microelectronics research, various nanostructure fabrication, applications of electronic devices in space environments, and so on.

**ACKNOWLEDGMENTS**

This work is supported by the National Natural Science Foundation of China (Grant No. 12135002) and Open Fund of National Key Laboratory of Intense Pulsed Radiation Simulation and Effect (NKLIPR2321).